# QSearchNet: A Quantum Walk Search Framework for Link Prediction

Priyank Dubey
Texas A& M University
College Station, Texas, USA
priyank_dubey@tamu.edu

## Abstract

Link prediction is one of the fundamental problems in graph theory, critical for understanding and forecasting the evolution of complex systems like social and biological networks. While classical heuristics capture certain aspects of graph topology, they often struggle to optimally integrate local and global structural information or adapt to complex dependencies. Quantum computing offers a powerful alternative by leveraging superposition for simultaneous multi-path exploration and interference-driven integration of both local and global graph features. In this work, we introduce **QSearchNet**, a quantum-inspired framework based on Discrete-Time Quantum Walk (DTQW) dynamics and Grover's amplitude amplification. QSearchNet simulates a topology-aware quantum evolution to propagate amplitudes across multiple nodes simultaneously. By aligning interference patterns through quantum reflection and oracle-like phase-flip operation, it adaptively prioritizes multi-hop dependencies and amplifies structurally relevant paths corresponding to potential connections. Experiments on diverse real-world networks demonstrate competitive performance, particularly with hard negative samples under realistic evaluation conditions.

## 1 Introduction

As the foundation for interpreting how networks expand and evolve, link prediction is one of the fundamental problems in network research and graph theory. Predicting future links or missing connections between nodes in a network using topological structures is a crucial topic of research [1][2]. This concept has broad and impactful applications in various domains, including social networks [3], biological systems [4], recommendation engines [5], and drug discovery [6]. Despite its broad utility, the task remains inherently challenging due to the complexity and diversity of real-world networks.

Classical link prediction has long relied on traditional heuristic methods that exploit topological patterns, broadly categorized into local and global approaches. Local neighbor-based heuristics, such as Common Neighbors (CN) [7], Adamic-Adar (AA) [8], and Resource Allocation (RA) [9], quantify link likelihood through direct neighborhood overlaps (e.g., shared neighbors) or degree-weighted interactions. While computationally efficient, these methods fail to capture dependencies beyond immediate neighbors, rendering them inefficient in sparse or hierarchically structured networks. Conversely, global path-based heuristics like the Katz Index [10] aggregate all paths between nodes, exponentially damping contributions by path length, while Shortest Path metrics [11] prioritize minimal geodesic distances. Though more holistic, global methods amplify noise in long-range paths and scale poorly to large graphs due to dense matrix operations. Critically, neither category adaptively integrates local and global signals, and both rely on rigid, handcrafted rules that struggle with "hard negatives"—topologically ambiguous node pairs that mimic genuine connections [12]. This limits their robustness in real-world applications, from social networks with latent communities to biological systems where functional links transcend proximity.

Quantum computing has emerged as a multidisciplinary computing paradigm that utilizes quantum mechanics to solve complex problems. The use of quantum states to encode the topological properties of graphs has already been investigated both theoretically [13] [14] [15] and experimentally [16]. Several quantum-enhanced techniques have been proposed for graph-related problems, including shortest path [17][18][19] and graph traversal [20][21], especially in highly connected networks. Recent advances include quantum-enhanced graph neural networks [22][23][24][25] and the quantum approximation optimization algorithm (QAOA) [26][27][28]. A quantum link prediction algorithm (QLP), based on continuous-time quantum walks (CTQW), has been proposed for link prediction [29][30]. It encodes the path-based prediction scores through the Hamiltonian derived from the adjacency matrix of the graph. QLP strictly depend on the d-sparse matrix model assuming homogeneous sparsity (i.e., largest node degree); this assumption reduces the method's practical usages over real-life networks, which actually show highly connected hubs along with a majority of sparsely connected nodes.

A promising alternative to CTQW is the discrete-time quantum walk(DTQW), which gives more precise control by allowing finer tuning of transition operators at each step of the quantum walk. Due to its granular resolution, it allows for tailored exploration in network structures; thus, DTQW could be well suited for complex or heterogeneous networks. Also, DTQW when coupled with Grover's algorithm [31], known for a quadratic speedup when searching unsorted databases, provides the foundation for quantum walk search algorithms explored in [32][33][34][35][36]. This combination uses the strengths of both methods, Grover's amplitude amplification for the relevant paths, and DTQW's graph exploration capability.

Quantum computing has gone through significant development in the last few years; to the best of our knowledge, there is no quantum walk search algorithm designed for the link prediction problem. This motivates us to explore the link prediction from the perspective of DTQW integrated with Grover's algorithm. In this work, we present **QSearchNet**- **Q**uantum **Search Net**work, a novel link prediction framework based on DTQW dynamics and Grover-inspired amplitude amplification. It integrates the principles of Grover's algorithm with the quantum projection and reflection-like operations to identify potential connections while simultaneously exploring multiple paths within the network. Our theoretical analysis demonstrates that QSearchNet unifies classical heuristics under specific parameterization, and it achieves exponential noise suppression, providing a fundamental advantage over traditional methods. In order to assess its efficacy, we performed extensive experiments on standard graph datasets and compared it's performance with standard heuristics. Furthermore, we evaluated QSearchNet under a more realistic evaluation condition using the Heuristic-Related Sampling Technique (HeaRT) [12], which uses various heuristics to sample hard negative samples. Our results show that the proposed framework consistently outperforms all traditional heuristics in HeaRT settings while performing competitively in a normal evaluation setting.

This paper is organized as follows. In Section 2, we briefly discuss the relevant quantum backgrounds ( DTQW and Grover's Algorithm). Section 3 provides the mathematical formulation and theoretical analysis of the proposed framework. In Section 4, we perform experiments and present the results in normal and HeaRT evaluation conditions. We also conduct ablation studies in this section. Finally, we provide the conclusion in Section 5.

## 2 Quantum Backgrounds

In this section, we present the concept of the DTQW and Grover's Algorithm, which are the key components of our research.

### 2.1 Discrete-Time Quantum Walk (DTQW)

Quantum walk is the quantum equivalent of classical random walks, crucial for designing quantum algorithms [37]. In general, there are two types of quantum walk: discrete-time quantum walk (DTQW), which is the focus of our work, and continuous-time quantum walk (CTQW) [38] [39]. Unlike their classical counterparts, quantum walks use quantum superposition and quantum interference instead of stochastic transitions to evolve over a graph in a more complex and efficient manner [40][41]. DTQW work in discrete time steps and are a key in quantum computing tasks such as search algorithms and distinguishing elements [40][42].

The generic form of any DTQW can be defined as the repeated application of a walk evolution operator $U_W$ on a composite Hilbert space, which is the tensor product of the position space augmented by either a coin space [21] or an auxiliary space of the same dimension as the original position space [43]. Mathematically, the generic DTQW evolution after $k$ steps can be expressed as:

$$|\psi_k\rangle = U_W^k |\psi_0\rangle,$$

Depending on the usage of coin space, DTQW can be divided into coined or coinless [44][34]. Coinless quantum walks have gained popularity due to their small Hilbert space and simple evolution operators, which can be obtained purely from graph characteristics. These walks have been shown to be as efficient as coined walks when used as a quantum search algorithm [32] [33][34][35]. In this work, we use a coinless DTQW in order to simulate the quantum walk dynamics for graph exploration.

### 2.2 Grover's Algorithm

Grover's algorithm is a quantum search algorithm presented by Grover in 1997 [31]. It demonstrates a quadratic speedup for unstructured search problems compared to classical algorithms. Given an unsorted database of $N$ elements, the task is to find a marked element (or verify its absence) with high probability. Classically, this problem would require $O(N)$ queries in the worst case, whereas Grover's algorithm achieves this with only $O(\sqrt{N})$ queries. It does this by iteratively applying two key components: the oracle operator and the diffusion operator. The oracle encodes the solution of the problem by flipping the phase of the marked element, defined as:

$$O|x\rangle = \begin{cases} -|x\rangle & \text{if } x \text{ is a solution,} \\ |x\rangle & \text{otherwise.} \end{cases}$$

This phase flip distinguishes the marked element from the rest. The diffusion operator amplifies the probability amplitude of the marked state while reducing the amplitudes of the non-marked states. It is sometimes referred to as the inversion about the mean operator and is defined as:

$$D = 2|\psi\rangle\langle\psi| - I$$

where, $|\psi\rangle$ is the equal superposition state of all elements. The amplitude amplification technique is the critical part of Grover's algorithm, which increases the probability of observing the desired state. In [32], authors combined the coinless DTQW along with the Grover's algorithm to design an algorithm for quantum search on the spatial grid.

## 3 Proposed Framework

In this section, we present the proposed QSearchNet framework for link prediction. Our approach is inspired by DTQW and Grover's algorithm, which are briefly discussed in Sections 2.1 and 2.2, respectively. The following subsections detail problem formulation

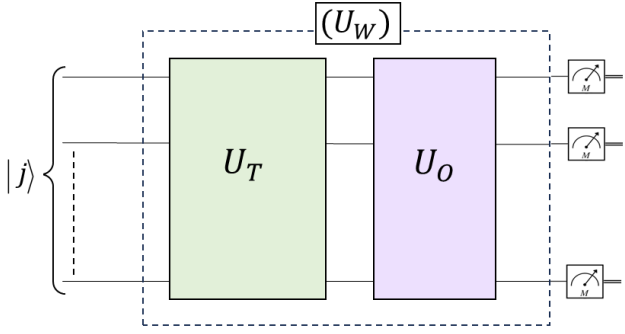

**Figure 1: Quantum circuit implementation of QSearchNet's walk evolution operator $U_W$. The register is initialized to $|j\rangle$, followed by application of the transition operator $U_T$ and oracle operator $U_O$. The dashed enclosure represents $U_W = U_O \cdot U_T$, which combines reflection about the graph topology ($U_T$) and target-node phase inversion ($U_O$). Qubit count scales as $\lceil \log_2 N \rceil$ for $N$-node graphs.**

& initialization, quantum walk dynamics, measurement and probability calculations.

## 3.1 Problem Formulation & Initialization

Consider a graph $G(V, E)$, where $V$ represents the set of vertices and $E$ represents the set of edges. Every edge $e \in E$ indicates a connection between nodes $u, v \in V$, forming a pair $(u, v)$. To mathematically represent the structure of the graph, we define the adjacency matrix $A$, where:

$$A^{u,v} = \begin{cases} 1 & \text{if } (u, v) \in E \\ 0 & \text{otherwise} \end{cases} \quad (1)$$

The adjacency matrix $A$ encodes the graph connectivity pattern, with $A^{u,v} = 1$ indicating the presence of an edge between the nodes $u$ and $v$. We can also express the adjacency matrix in Dirac notation as:

$$A = \sum_{i,j \in E} A_{ij} |i\rangle\langle j| \quad (2)$$

Furthermore, the degree $d_i$ of node $i$ is defined as the sum of the elements in the corresponding row of the adjacency matrix:

$$d_i = \sum_{j=1}^{N} A_{ij} \quad (3)$$

This counts the number of connections that lead to (or away from) node $i$. Using $d_i$ as diagonal elements, we define degree matrix $D$ as:

$$D = \sum_{i=1}^{N} d_i |i\rangle\langle i| \quad (4)$$

For a graph with $N$ nodes, each node $v \in V$ can be represented as the basis state $|v\rangle$ in an $N$ dimensional Hilbert space, where each basis state corresponds to a localized state at a specific node of the graph. The quantum state represents the walker's position on the graph, and the number of qubits required to represent this depends on the number of nodes in a graph. Specifically, for the graph with $N$ nodes, $\lceil \log_2 N \rceil$ qubits are required. Thus, QSearchNet requires $\lceil \log_2 N \rceil$ qubits in total. The walk register is initialized at a starting node $|j\rangle$, and the initial quantum state is given by:

$$|\psi_0\rangle = |j\rangle_n, \quad (5)$$

where $|j\rangle$ denotes the initial localized state at node $j$. For $N$-node graph, it can be written as $|\psi_0\rangle = [0, 0, \ldots, 1, \ldots, 0]^T$, where 1 is at the $j$-th position, while other entries remain 0. This initialization ensures that the quantum walker begins its exploration from a specific node in the graph.

## 3.2 Quantum Walk Dynamics

This section outlines the core component of our framework. The key operators of the Quantum Walk include the Transition Operator ($U_T$), Oracle Operator ($U_T$), and Walk Evolution Operator ($U_W$).

*3.2.1 Transition Operator ($U_T$).* The transition operator controls the walker's movement across the graph nodes by manipulating the quantum state based on the topological properties. To define $U_T$, we first introduce the transition probabilities subspace $(P_T)^1$, based on hermitian quantum walk generator. This is derived from the symmetric normalized adjacency matrix (or supra-adjacency matrix) as:

$$P_T = D^{-1/2} A D^{-1/2} = \sum_{i,j} P(i,j) |i\rangle\langle j| \quad (6)$$

where, $A$ is adjacency matrix, $D$ is degree matrix, and $P(i, j)$ represents transition probabilities between the nodes $i$ and $j$ in the graph. The action of $P_T$ is similar to projection operators in quantum mechanics. It acts on the Hilbert space of the walker and confines its evolution to the subspace defined by these transition probabilities. For an initial quantum state $|\psi_0\rangle$, localized at the starting node $|j\rangle$, $P_T$ redistributes the walker's amplitude among neighboring nodes according to transition probabilities:

$$|\psi_1\rangle = P_T |\psi_0\rangle = \sum_{i,j} P(i,j) \langle j|\psi_0\rangle |i\rangle \quad (7)$$

The subspace defined by $P_T$ provides a basis for representing both local and global structural properties of the graph in quantum state space. In order to go beyond classical-like redistribution and utilize quantum advantages, we define transition operator as:

$$U_T = (2P_T - I) \quad (8)$$

This operator reflects the quantum state about the subspace defined by the graph topology ($P_T$). The transition operator $U_T$ allows constructive interference for paths closely aligned with the initial node and destructive interference for paths less relevant. This dynamics enhances the quantum walk's ability to emphasize meaningful structural patterns in the graph. It intrinsically balances the contribution of local adjacency relationships with the global spectral structure of the graph.

---

[1] The operator $P_T$ is hermitian by construction, ensuring spectral alignment with the graph. However, it is unitary only specific graphs. See Appendix A for implementation of non-unitary Operators.

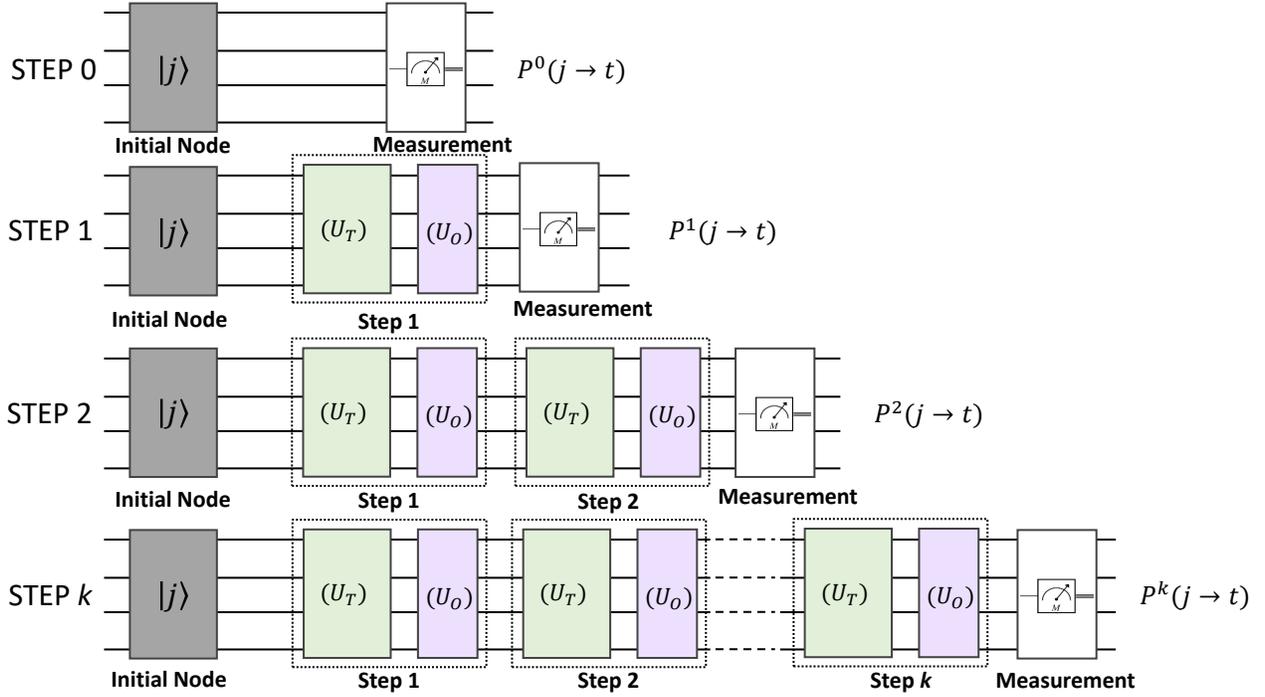

**Figure 2: Iterative application of $U_W = U_O \cdot U_T$ for $k$ walk steps. From top to bottom: Circuit diagrams for $k = 0$ (initial state $|j\rangle$), $k = 1$ ($U_W^1$), $k = 2$ ($U_W^2$), and general $k$-step evolution ($U_W^k$) with $P^k(j \to t)$ representing corresponding probability.**

*3.2.2 Oracle Operator ($U_O$).* The oracle operator is used to mark the target node $t$ by inverting its amplitude. The oracle operator $U_O$ is defined as:

$$U_O = (I - 2|t\rangle\langle t|) \tag{9}$$

where, $I$ is the identity operator on the walk register. $|t\rangle\langle t|$ is the projector on the target node $t$. The term $(I - 2|t\rangle\langle t|)$ inverts the amplitude of the target node $t$. Specifically, it transforms $|t\rangle$ to $-|t\rangle$ while leaving all other nodes unchanged. This mechanism selectively amplifies the amplitude at the target node $t$ through phase inversion, similar to Grover's amplitude amplification as discussed in Section 2.2. This shifts the interference pattern in favor of the paths leading to the target node. After applying $U_O$, the quantum state evolves as:

$$|\psi_2\rangle = U_O|\psi_1\rangle \tag{10}$$

or equivalently,

$$|\psi_2\rangle = U_O \cdot U_T|\psi_0\rangle, \tag{11}$$

*3.2.3 Walk Evolution Operator ($U_W$).* The Walk Evolution Operator ($U_W$) governs the overall evolution of the quantum state by iteratively applying both the Transition Operator ($U_T$) and the Oracle Operator ($U_O$) as shown in Figure 1. It controls the evolution of the quantum state with each step of the walk. The Walk Evolution Operator for a single walk step is as follows:

$$U_W = U_O \cdot U_T \tag{12}$$

where: $U_T$ reflects the quantum state $|j\rangle$ onto the subspace $P_T$, effectively inverting components of the state orthogonal to the defined subspace while amplifying those aligned with it. $U_O$ flips the phase at the target node $t$, ensuring that interference patterns favor the target node during subsequent steps. This synergistic action enables the quantum walk to systematically emphasize significant patterns in the network, increasing link prediction efficiency. The process of applying $U_W$ is iterated for a predefined number of steps (e.g., $k$ steps) as shown in Figure 2. At each step, the quantum state evolves as:

$$|\psi_{k+1}\rangle = U_W|\psi_k\rangle = U_O U_T|\psi_k\rangle), \tag{13}$$

where, $|\psi_k\rangle$ is the quantum state after $k$ steps of the walk.

### 3.3 Measurement and Probability Calculations

The measurement and probability calculation are essential for converting the abstract amplitude distributions of the quantum walk into quantifiable metric for link prediction. The measurement procedure involves directly measuring the walk register after the quantum walk evolution. After completing $k$ steps of the quantum walk, the walker's state is represented as:

$$|\psi_k\rangle = U_W^k|j\rangle_n, \tag{14}$$

where, $|j\rangle_n$ represents the quantum state corresponding to the starting node $j$. After the walk evolution, the walk register is measured, collapsing the quantum state to a specific target node $\langle t|$ with a

**Table 1: Benchmark datasets statistics, including the number of nodes, edges, average density and split ratio for train/validation/test.**

|  | Cora | Citeseer | Pubmed | ogbl-collab | ogbl-ddi |
| --- | --- | --- | --- | --- | --- |
| Nodes | 2,708 | 3,327 | 18,717 | 235,868 | 4,267 |
| Edges | 5,278 | 4,676 | 44,327 | 1,285,465 | 1,334,889 |
| Avg. degree | 3.90 | 2.81 | 4.74 | 10.90 | 625.68 |
| Split Ratio | 85 / 5 / 10 | 85 / 5 / 10 | 85 / 5 / 10 | 92 / 4 / 4 | 80 / 10 / 10 |

probability proportional to:

$$P(t) \propto |\langle t|U_W^k|j\rangle_n|^2 \quad (15)$$

where, $P(t)$ represents the likelihood of the walker being at node $t$ after $k$ steps. A high value of $P(t)$ suggest strong structural relationships between the nodes $j$ and $t$, such as shared neighbors, positively weighted paths, or other structural features of the graph that indicate a possible link. Conversely, a low value of $P(t)$ suggests weak or limited relationships, reducing confidence in the existence of a possible link.

In summary, for each initial node $j$, the algorithm evaluates potential connections to node $t$ using quantum walk dynamics. The walk evolution operator $U_W = U_O \cdot U_T$ governs this process, where $U_T$ represents the transition operator that redistributes the amplitude across the graphs by preferring the relevant connections of the initial node, and $U_O$ is the oracle operator that selectively marks the target node $t$ by inverting its amplitude. At each step $k$, the quantum state evolves as $|\psi_{k+1}\rangle = U_W|\psi_k\rangle$, leveraging interference to amplify connections aligned with topological patterns, prioritizing paths between the initial and target nodes. After the quantum walk is completed, the probability $P(t) \propto |\langle t|U_W^k|j\rangle_n|^2$ is calculated, representing the likelihood of link for each pair $(j, t)$.

### 3.4 Theoretical Analysis

We analyze the theoretical foundations of QSearchNet through both spectral analysis of quantum walk dynamics and a path-integral interpretation. A detailed discussion of these individual analysis are provided in Appendices C.1 and C.2. From equations 6 and 8, the transition operator $U_T = 2P_T - I$, with $P_T = \sum_{i,j} P(i,j)|i\rangle\langle j|$, governs the propagation of the amplitude. For undirected graph, $P_T$ is hermitian and its spectral decomposition is given by:

$$P_T = \sum_{i=1}^{N} \lambda_i |v_i\rangle\langle v_i|, \quad (16)$$

so that,

$$U_T = \sum_{i=1}^{N} (2\lambda_i - 1)|v_i\rangle\langle v_i| = \sum_{i=1}^{N} \mu_i |v_i\rangle\langle v_i| \quad (17)$$

where $\lambda_i$ are eigenvalues of $P_T$ and $|v_i\rangle$ its eigenvectors. The eigenvalues $\mu_i = 2\lambda_i - 1$ of $U_T$ determine the evolution of the amplitude. The iterative process of applying overall walk evolution operator $U_W$ generates time-dependent coefficients $c_i^{(k)}$, such that state $|\psi_k\rangle$ is given as:

$$|\psi_k\rangle = \sum_{i=1}^{N} c_i^{(k)} |v_i\rangle \quad (18)$$

with $c_i^{(k)}$ depending on both $\mu_i^k$ and the cumulative phase inversions introduced by the oracle operator $U_O$. From the path integral perspective, the probability of reaching a target node $t$ in $k$ steps is given by $P(t) = |\langle t|U_W^k|j\rangle|^2$. This probability arises from coherently summing amplitudes over all $k$-step paths from $j$ to $t$:

$$\langle t|U_W^k|j\rangle = \sum_{\substack{\text{paths } p: j \to t \\ \text{in } k \text{ steps}}} (-1)^{n_p} \prod_{\ell=1}^{k} \left[ \frac{2A_{u_\ell u_{\ell+1}}}{\sqrt{d_{u_\ell} d_{u_{\ell+1}}}} - \delta_{u_\ell, u_{\ell+1}} \right] \quad (19)$$

where, the product over $\ell$ accumulates per-step amplitudes during propagation. $\frac{2A_{A_\ell+\ell+1}}{\sqrt{d_\psi/d_{\psi+1}}}$ represents the amplitude for a transition from node $u_l$ to $u_{l+1}$. The subtraction term $-\delta_{u_\ell,u_{\ell+1}}$ arises from $-I$ term of the $U_T$. It flips the sign of the component orthogonal to the subspace defined by $P_T$, ensuring that the state is reflected rather than simply allowing for a self-loop. The factor $(-1)^{n_p}$ accounts for the oracle-induced phase inversions, where $n_p$ is the number of times the target node $|t\rangle$ is encountered along the path $p$. Our analysis establishes two key properties of QSearchNet:

**Theorem 3.1** (Unification of Classical Heuristics). *QSearchNet's score $P(t)$ subsumes classical heuristics methods Common Neighbors (CN), Resource Allocation (RA), Adamic-Adar (AA), and Katz Index under specific parameterizations. (Proof: Appendix C.3.)*

**Theorem 3.2** (Exponential Noise Suppression). *QSearchNet suppresses noise or topologically irrelevant paths between the initial node $j$ and target node $t$ at $O(e^{-4\Delta k})$, outperforming classical methods by an exponential margin. (Proof: Appendix C.4.)*

## 4 Experiments

In this section, we conduct experiments on standard benchmark datasets to validate the effectiveness of the proposed QSearchNet. The classical simulation used to calculate link prediction scores on a conventional computer is described in the Appendix D and is available on Anonymous GitHub [2].

### 4.1 Datasets and Experimental Settings

*4.1.1 Datasets.* To evaluate our proposed framework, we used four commonly used attributed graphs for link prediction. We used small-scale datasets like Cora, Citeseer, and Pubmed [45], as well as a subset of large-scale datasets from the OGB benchmark [46], such as ogbl-collab and ogbl-ddi. Table 1 provides the statistics for each dataset. For additional detail for experimental settings, refer to the Appendix B.

---
[2]https://anonymous.4open.science/r/QSearchNet/

Table 2: Results on benchmark datasets (%) under normal evaluation setting. Highlighted are the results ranked first, second, and third.

| Category | Model | Cora (MRR) | Citeseer (MRR) | Pubmed (MRR) | ogbl-collab (Hits@50) |
|---|---|---|---|---|---|
| Heuristic | CN | 20.99 | 28.34 | 14.02 | 61.37 |
|  | AA | 31.87 | 29.37 | 16.66 | 64.17 |
|  | RA | 30.79 | 27.61 | 15.63 | 63.81 |
|  | Shortest Path | 12.45 | 31.82 | 7.15 | 46.49 |
|  | Katz | 27.40 | 38.16 | 21.44 | 64.33 |
| Quantum Algorithm | QSearchNet | 30.15 | 40.28 | 39.43 | 61.09 |

Table 3: Hyperparameter values used for each dataset.

| Data set | Number of Steps ($k$) |
|---|---|
| Cora | 2 |
| Citeseer | 4 |
| Pubmed | 3 |
| ogbl-collab | 2 |
| ogbl-ddi | 2 |

*4.1.2 Hyperparameters.* For each dataset, we selected the number of walk steps ($k$) that maximize the prediction score, as presented in Table 3.

*4.1.3 Baseline Models.* QSearchNet is evaluated against classical heuristic methods, spanning two categories of topological analysis: (1) **Local neighbor-based approaches**, including Common Neighbors (CN) [7], Adamic-Adar (AA) [8], and Resource Allocation (RA) [9], which predict links using direct neighborhood structures such as shared neighbors and degree-weighted interactions; and (2) **Global path-based methods**, such as Katz Index [10], which aggregates all paths between nodes with exponential damping, and Shortest Path [11], relying on minimal path lengths. The corresponding results are taken from [12]. These baselines represent foundational strategies in link prediction, from granular local interactions to holistic path analysis, enabling a rigorous assessment of QSearchNet's ability to unify structural insights across multiple scales.

*4.1.4 Evaluation Metrics.* For evaluation, we employ ranking-based metrics, specifically mean reciprocal rank (MRR) and Hits@K, described in Appendix B.1. For the OGB dataset, we adopt the metrics used in their original studies: Hits@50 for ogbl-collab. For the small-scale datasets Cora, Citeseer, and Pubmed, we use MRR as the evaluation metric.

### 4.2 Results

The results of our experiments are summarized in Table 2. We observe that under normal evaluation setting, QSearchNet demonstrates competitive performance on all three small-scale datasets. For Citeseer and Pubmed datasets, it outperforms all baseline models, showcasing its ability to effectively capture the structural intricacies of these graphs. Specifically, QSearchNet outperforms the best classical heuristic by approximately 84 % on Pubmed dataset. For Cora, it performs comparably to top classical heuristics such as Katz and Resource Allocation, indicating robust performance even in datasets where simpler heuristics are traditionally strong. However, its performance on the large ogbl-collab dataset is only average compared to the other baseline models. ogbl-ddi is excluded from the main results in the normal evaluation setting due to recent issues identified in [12] (refer to Appendix B.4 for more details).

It can be inferred from the results obtained on small datasets that QSearchNet is capable of capturing rich structural patterns in graph. The iterative quantum walk efficiently propagates the amplitude across multiple nodes simultaneously, while highlighting meaningful patterns with constructive interference at the target node. The average performance on the ogbl-collab dataset could be due to the inherently complex and sparse nature of the structure in this dataset. Compared to small-scale datasets, in which rich and dense structural information could be fully mined, large-scale graphs often suffer from higher-level noise with a less distinctive pattern of connectivity, hard to capture by quantum dynamics.

### 4.3 Performance of QSearchNet under HeaRT Evaluation

To further assess QSearchNet, we further evaluate its performance using the HeaRT evaluation framework [12]. This framework introduces a more realistic and challenging setting for link prediction by generating harder negative samples through heuristic-based techniques. Unlike the existing evaluation setting that uses randomly sampled negatives—often trivial to classify—HeaRT personalizes the negative samples, resulting in a significantly more difficult prediction task. The detailed results with HeaRT evaluation are provided in Table 4. QSearchNet outperforms all classical heuristics across all datasets, including the previously excluded ogbl-ddi and the challenging large-scale ogbl-collab, achieving top rank in every benchmark datasets.

Interestingly, QSearchNet achieves state-of-the-art (SOTA) performance on the ogbl-collab dataset in the HeaRT framework outperforming all baseline models including graph neural networks (GNNs) and embedding methods [12], which is a significant improvement over its average results in standard evaluation. This improved performance indicates that algorithm benefits from the more realistic negative samples where global heuristics like Katz Index fail to distinguish meaningful long-range collaborations from noisy connections, while local methods like Resource Allocation (RA) over-penalize low-degree nodes in sparse networks. QSearchNet, however, overcomes these challenges through its quantum-inspired methods to balance localized collaboration motifs (e.g., co-author triads) with signals from intermediary hubs, while interference-driven reflection amplifies coherent multi-hop paths. This enables a 7.21% MRR under HeaRT—surpassing Katz (6.31%) and marking a 14% improvement over its own average performance in standard evaluation. Similarly, on ogbl-ddi, where classical heuristics like the Shortest Path collapse entirely due to disconnections. QSearchNet's phase-inversion mechanism suppresses noise from synthetic irregularities while amplifying latent biochemical dependencies (e.g., indirect protein-mediated interactions), achieving 10.53% MRR.

Table 4: Results on benchmark datasets (%) under HeaRT evaluation. Highlighted are the results ranked first, second, and third.

| Category | Model | Cora (MRR) | Citeseer (MRR) | Pubmed (MRR) | ogbl-collab (MRR) | ogbl-ddi (MRR) |
|---|---|---|---|---|---|---|
| Heuristic | CN | 9.78 | 8.42 | 2.28 | 4.20 | 6.71 |
| | AA | 11.91 | 10.82 | 2.63 | 5.07 | 6.97 |
| | RA | 11.81 | 10.84 | 2.47 | 6.29 | 8.70 |
| | Shortest Path | 5.04 | 5.83 | 0.86 | 2.66 | 0 |
| | Katz | 11.41 | 11.19 | 3.01 | 6.31 | 6.71 |
| Quantum Algorithm | QSearchNet | 13.01 | 15.89 | 8.11 | 7.21 | 10.53 |

Table 5: MRR (%) With vs. Without Oracle Operator.

| Data set | MRR (With Oracle) | MRR (With Oracle) |
|---|---|---|
| Cora | 30.15 | 26.19 |
| Citeseer | 40.28 | 40.28 |
| Pubmed | 39.43 | 38.37 |

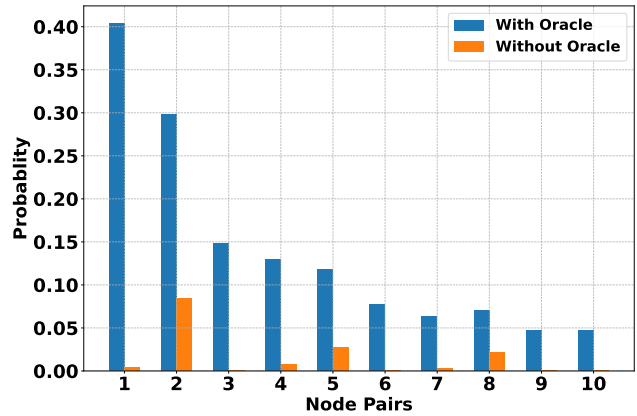

### 4.4 Ablation Studies

We conduct ablation studies to systematically evaluate the contribution of key components in QSearchNet, focusing on small-scale datasets (Cora, Citeseer, Pubmed). First, we analyze the impact of the depth of the quantum walk by varying the number of steps $k$ from 1 to 10, probing how the propagation range influences the extraction of structural patterns. Second, we remove the oracle operator $U_O$ (responsible for the inversion of the target node phase) to isolate the role of interference-driven amplification.

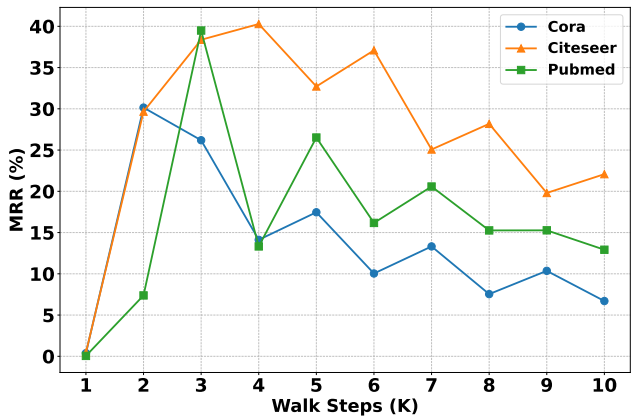

Figure 3: Mean Reciprocal Rank (MRR) vs. Quantum Walk Steps (k) for Cora, Citeseer, and Pubmed. Performance peaks at $k = 2$ for Cora (30.15%), $k = 4$ for Citeseer (40.28%), and $k = 3$ for Pubmed (39.49%), reflecting dataset-specific structural depths.

Figure 4: Amplitude probabilities at target nodes with and without oracle. Target-node probabilities drop by orders of magnitude when the oracle is removed, illustrating its role in constructive interference.

The optimal quantum walk depth ($k$) varies across datasets, reflecting their structural properties (Fig. 3). For Cora, performance peaks at $k = 2$ (MRR: 30.15%), as its dense, community-driven triads require only short-range walks to capture direct co-citation patterns. Longer steps ($k > 2$) propagate noise into overlapping clusters, reducing precision. In contrast, Citeseer achieves the maximal MRR (40.28%) at $k = 4$, which requires a deeper exploration to resolve the interdisciplinary citation chains in its sparse network. For Pubmed, hierarchical dependencies balance at $k = 3$ (MRR: 39.49%), while $k > 3$ disperses amplitudes into tangential references, degrading performance. These trends underscore the importance of adaptive walk-step tuning: shallow steps fit sparse graphs (e.g., Citeseer's fragmented topology), while excessive steps overfit dense ones (e.g., Cora's overlapping communities) by amplifying noise.

Removing the oracle operator ($U_O = I - 2|t\rangle\langle t|$) reveals its critical role in amplifying latent connections demonstrated through its impact on both amplitude distributions (Fig. 4) and MRR metrics (Table 5). For Cora, the target node probabilities collapse without $U_O$ dropping orders of magnitude (40.4% to 0.37% for some pairs), while MRR drops 13.1% (30.15% to 26.19%), as phase inversion no longer suppresses noise from competing triads. In Pubmed, hierarchical dependencies exhibit similar sensitivity, although the overall decline in MRR is smaller (39.43% to 38.37%), reflecting partial self-organization of the citation chains. In contrast, Citeseer shows no change in MRR (40.28% to 40.28%), as its sparse and structured topology lacks the coherent paths necessary for interference-driven amplification. Here, classical-like propagation suffices, and the inversion of the oracle phase becomes redundant.

In conclusion, ablation studies validate two core components of QSearchNet: adaptive walk-step tuning and interference-driven amplification. The optimization of steps aligns with structural diversity, where shallow steps ($k = 2$) prevent noise amplification in Cora's dense communities, while deeper steps ($k = 4$) model the sparse interdisciplinary links of Citeseer, and intermediate steps ($k = 3$) balance Pubmed's hierarchical dependencies. The oracle operator $U_O$, critical in noise-rich (Cora) or hierarchical (Pubmed) regimes, suppresses ambiguities via destructive and constructive interference, but becomes redundant in sparse topologies (Citeseer) where classical propagation dominates. By integrating classical path aggregation with quantum-inspired mechanics, QSearchNet addresses the inability of classical heuristics to resolve edge cases in complex graphs.

## 5 Conclusion

We introduced QSearchNet, a quantum-inspired link prediction framework that integrates discrete-time quantum walk with Grover-inspired amplitude amplification. By simulating multi-path exploration with quantum walk and leveraging phase inversion, QSearchNet dynamically balances the local and global structural contributions. Theoretically, QSearchNet subsumes classical heuristics under specific parameterization, while its interference-based amplification mechanism achieves exponential noise suppression compared to traditional methods. Experiments show that QSearchNet achieves competitive performance across multiple standard datasets, particularly under the HeaRT evaluation setting. This work establishes the groundwork for applying Grover-like amplification to graph-based link prediction tasks. In future work, we plan to analyze the complexity of QSearchNet and its further application on heterogeneous and dynamic graphs.

## A Implementation of Non-Unitary Operators

Recent developments in quantum computing have emphasized the importance of non-unitary operations like $P_T$, which are essential for simulating real-world quantum phenomena, from open systems with energy dissipation to Quantum machine learning[47]. These operators go beyond the limitations of unitary frameworks, enabling algorithms to model noise, loss, and topological phases intrinsic to biological, chemical, and social networks. This work prioritizes the validation of the quantum-inspired advantages of QSearchNet. The specific implementation details of the operator are beyond the scope of this work. However, the growing research interest in nonunitary quantum dynamics has motivated the development of several frameworks to implement such operators, which are critical for simulating real-world phenomena. Examples include Biorthogonal dilation (unitary embedding via eigenstate biorthogonality) [48], Linear combinations of unitaries (LCU) (probabilistic simulation with phase-matching optimizations) [49], Block encodings (ancilla-mediated unitary extensions) and Symmetry-Aware Normalization [50]. These methods have been proven to achieve scalable, noise-resilient implementations of nonunitary dynamics.

## B Additional detail on Experimental Settings

### B.1 Evaluation Metrics

Given $N$ positive samples where each positive sample is ranked among $M$ negative counterparts, let $\text{rank}_i$ denote the position of the $i$-th positive sample when sorted by prediction scores in descending order. The evaluation metrics are defined as:

- **Hits@K**:

$$\text{Hits@K} = \frac{1}{N} \sum_{i=1}^{N} \mathbb{I}(\text{rank}_i \leq K)$$

where $\mathbb{I}(\cdot)$ is the indicator function:

$$\mathbb{I}(\text{rank}_i \leq K) = \begin{cases} 1 & \text{if } \text{rank}_i \leq K \\ 0 & \text{otherwise} \end{cases}$$

- **Mean Reciprocal Rank (MRR)**:

$$\text{MRR} = \frac{1}{N} \sum_{i=1}^{N} \frac{1}{\text{rank}_i}$$

### B.2 Negative Sampling & Evaluation

QSearchNet, as a heuristic-based framework, does not require negative samples for training but adheres to established evaluation protocols and data-splits [12] for fair benchmarking. Negative samples $(a^-, b^-)$ are generated as non-existent edges through uniform random node pairing:

$$(a^-, b^-) = (\text{Rand}(V), \text{Rand}(V)), \tag{20}$$

where Rand(V) selects a random node from the graph $G(V, E)$. During evaluation, a fixed set of negative samples is used for each positive sample to ensure consistency. For a positive sample $(u, a)$, the negative samples are restricted to contain one of its two nodes through corruption. Formally, the negative samples belong to the set:

$$S(u, a) = \big\{(x, a) \mid x \in \mathcal{V}\big\} \cup \big\{(u, y) \mid y \in \mathcal{V}\big\} \tag{21}$$

where $\mathcal{V}$ is the vertex set of graph $G$. This constitutes all possible corruptions of either: the first node $u$ while keeping $a$ fixed, or the second node $a$ while keeping $u$ fixed with $x, y$ being randomly selected nodes from $\mathcal{V}$. For the OGB datasets, we use the set of fixed negatives that are randomly chosen [46]. Also, for ogbl-collab we follow the protocol in [46] and include the validation edges in the training graph while testing. This setting is adopted in ogbl-collab in both the existing and the new evaluation setting.

### B.3 Computation Resources

For our experiments, we have used three high-capacity GPU resources: NVIDIA L4, NVIDIA GeForce GTX 1080, and Quadro RTX 4000.

### B.4 Omission of ogbl-ddi under the normal Evaluation

We exclude the results for ogbl-ddi in Table 2, when using the existing evaluation settings due to limitations identified by Li et al. [12]. Specifically, they observed a weak correlation between validation and test performance for ogbl-ddi, which affects the reliability of hyperparameter tuning. This issue is particularly relevant to QSearchNet, as it relies solely on structural information without leveraging additional node or edge features. Consequently, tuning hyperparameters on the validation set does not guarantee consistent or high test performance, often resulting in significantly lower accuracy than the reported results.

Due to these concerns, we believe that ogbl-ddi, in its current evaluation setting, is not well-suited for evaluation with QSearchNet. Therefore, we omit its results under the existing setting. However, this issue is resolved when ogbl-ddi is evaluated using the newly proposed HeaRT framework. Hence, we include the QSearchNet results in the HeaRT evaluation setting in Table 4 for completeness.

## C Additional Analysis of QSearchNet

### C.1 Spectral Analysis

The transition operator is defined as

$$U_T = 2P_T - I$$

where, $P_T$ is given by the symmetric normalized adjacency matrix

$$P_T = D^{-1/2}AD^{-1/2} = \sum_{i,j} P(i,j)|i\rangle\langle j|$$

governs the propagation of amplitude within the framework. For undirected graphs, $P_T$ is hermitian with spectral decomposition as:

$$P_T = \sum_{i=1}^{N} \lambda_i |v_i\rangle\langle v_i|,$$

where, $\lambda_i \in [-1, 1]$ are eigenvalues and $|v_i\rangle$ are the corresponding orthogonal eigenvectors. Consequently, the spectral decomposition of $U_T$ is:

$$U_T = 2\sum_{i=1}^{N} \lambda_i |v_i\rangle\langle v_i| - I = \sum_{i=1}^{N} (2\lambda_i - 1) |v_i\rangle\langle v_i|.$$

where, $\mu_i = 2\lambda_i - 1$ are eigenvalues of $U_T$. Notre that $U_T$ is unitary if $|\mu_i| = 1$ for all $i$, and $\lambda_i \in \{0, 1\}$. This restricts $U_T$ to graphs where $P_T$ is a projection matrix (special cases). For the general case, $P_T$ is not a projection, $U_T$ may not be strictly unitary, necessitating the use of techniques for implementing non-unitary operators as discussed in the earlier section. The overall walk evolution operator $U_W = U_O U_T$, where $U_O = I - 2|t\rangle\langle t|$ combines the transition dynamics with the oracle operation. After $k$ steps, the state becomes:

$$|\psi_k\rangle = \underbrace{U_O U_T \cdot U_O U_T \cdots U_O U_T}_{k \text{ times}} |j\rangle$$

Each application of $U_W = U_O U_T$ involves a transition in which $U_T$ redistributes the amplitudes through:

$$U_T = \sum_{i=1}^{N} \mu_i |v_i\rangle\langle v_i|$$

followed by an oracle operation where $U_O$ inverts the phase of target node $|t\rangle$, modifying the coefficients of all eigenstates $|v_i\rangle$ that overlap with $|t\rangle$. The iterative process generates time-dependent coefficients $c_i^{(k)}$, where:

$$|\psi_k\rangle = \sum_{i=1}^{N} c_i^{(k)} |v_i\rangle$$

with $c_i^{(k)}$ depending on both $\mu_i^k$ and the cumulative phase inversions from $U_O$. This selective inversion amplifies contributions from eigenstates $|v_i\rangle$ with significant overlap $\langle t|v_i\rangle$. This biases the walk toward the paths connecting $|j\rangle$ to $|t\rangle$

### C.2 Path Integral Interpretation

The probability $P(t) = \left|\langle t|U_W^k|j\rangle\right|^2$ can be interpreted as a quantum-mechanical sum on all possible paths connecting the starting node $|j\rangle$ to the target node $|t\rangle$ through the steps $k$ of the quantum walk. This formulation explicitly captures how local transitions and global topological dependencies are unified through quantum-mechanical interference. To derive this, we begin with the iterative application of the overall evolution operator of the walk $U_W^k = (U_O U_T)^k$, where the transition operator is defined as $U_T = 2P_T - I$ with $P_T = D^{-1/2}AD^{-1/2}$. $P_T$ propagates amplitudes to neighboring nodes with weights defined by the normalized adjacency matrix as:

$$\langle u|P_T|v\rangle = \frac{A_{uv}}{\sqrt{d_u d_v}}$$

where, $A_{uv}$ is the binary representation of the connection between nodes $u$ and $v$ and $d_u, d_v$ are their respective degrees. With this, $U_T$ is defined as:

$$\langle u|U_T|v\rangle = \frac{2A_{uv}}{\sqrt{d_u d_v}} - \delta_{uv}$$

where, the first term $\frac{2A_{uv}}{\sqrt{d_u d_v}}$ corresponds to move from node $v$ to node $u$ as defined by $P_T$ and the second term $\delta_{uv}$ (from $-I$) subtracts the amplitude corresponding to the component that does not change (i.e., when the walker remains in the same state). In the reflection picture, it flips the sign of the component orthogonal to the subspace defined by $P_T$, ensuring that the state is reflected rather than simply allowing for a self-loop. The oracle operator $U_O = I - 2|t\rangle\langle t|$ introduces a phase inversion at the target node $|t\rangle$. Every time the state $|t\rangle$ is encountered during evolution, this operator multiplies its amplitude by $-1$. After $k$ steps, the amplitude $\langle t|U_W^k|j\rangle$ can be represented as a coherent sum over all paths $p$ which take the walker from $j$ to $t$ in $k$ steps. If we denote the sequence of nodes along a given path by $p: u_1, u_2...u_{k+1}$, where $u_1 = j$ and $u_{k+1} = t$, then the amplitude is given by:

$$\langle t|U_W^k|j\rangle = \sum_{\substack{\text{paths } p; j \to t \\ \text{in } k \text{ steps}}} (-1)^{n_p} \prod_{\ell=1}^{k} \left[ \frac{2A_{u_\ell u_{\ell+1}}}{\sqrt{d_{u_\ell} d_{u_{\ell+1}}}} - \delta_{u_\ell, u_{\ell+1}} \right]$$

where, $\frac{2A_{A_4+t+1}}{\sqrt{d_\psi/d_{\psi+1}}}$ represents the amplitude for a transition from node $u_l$ to $u_{l+1}$. $-\delta_{u_\ell, u_{\ell+1}}$ is the subtraction component from $-I$ term of the $U_T$. The factor $(-1)^{n_p}$ accounts for the oracle-induced phase factor with $n_p$ representing the number of times the path visits the target node $t$. The interference mechanism in the quantum walk thus operates as follows:

- **Constructive Interference**: Paths that visit the target node $t$ with even $n_p$ (e.g., paths visiting $t$ 0,2,4, ... times) contribute positively ($(-1)^{n_p} = +1$), amplifying their collective weight.
- **Destructive Interference**: Paths that visit the target node $t$ with odd $n_p$ (e.g., paths visiting $t$ 1,3,5, ... times) contribute negatively ($(-1)^{n_p} = -1$), suppressing incoherent or noisy detours.

The final probability $\left|\langle t|U_W^k|j\rangle\right|^2$ computes the coherent sum of all amplitude contributions, where constructive/destructive interference occurs at the amplitude level prioritizing paths that are structurally significant for link prediction between nodes $j$ and $t$.

### C.3 Unification of Classical Heuristics

**Theorem C.1** (Unification). *Let $P(t) = |\langle t|U_W^k|j\rangle|^2$ be the link prediction score computed by QSearchNet, where $U_W = U_O U_T$, $U_T = 2P_T - I$, $P_T = D^{-1/2}AD^{-1/2}$, and $U_O = I - 2|t\rangle\langle t|$. Under specific parameterizations, $P(t)$ approximates classical heuristics as follows:*

- **Degree-Normalized Common Neighbors (CN)**: *For $k = 2$, $U_O = I$:*

$$P(t) \propto \left|\sum_k \frac{A_{j,k} A_{k,t}}{\sqrt{d_j d_t} d_k}\right|^2.$$

- **Resource Allocation (RA)**: *For $k = 2$, $U_O = I$, and reweighted adjacency matrix $\tilde{A}_{j,k} = \frac{A_{j,k}}{d_k}$:*

$$P(t) \propto \left|\sum_k \frac{\tilde{A}_{j,k} \tilde{A}_{k,t}}{\sqrt{\tilde{d}_j \tilde{d}_k \tilde{d}_t}}\right|^2 \approx \left(\sum_{k \in \Gamma(j) \cap \Gamma(t)} \frac{1}{d_k}\right)^2.$$

- **Adamic-Adar (AA)**: *For $k = 2$, $U_O = I$, and reweighted adjacency matrix $\tilde{A}_{j,k} = \frac{A_{j,k}}{\log d_k}$:*

$$P(t) \propto \left|\sum_k \frac{\tilde{A}_{j,k} \tilde{A}_{k,t}}{\sqrt{\tilde{d}_j \tilde{d}_k \tilde{d}_t}}\right|^2 \approx \left(\sum_{k \in \Gamma(j) \cap \Gamma(t)} \frac{1}{\log d_k}\right)^2,$$

*where $\tilde{d}_i = \sum_m \tilde{A}_{i,m}$.*

- **Spectral Katz Analogy**: *For $k \gg 1$, $U_O = I$:*

$$P(t) \propto \left|\sum_{i=1}^{N} (2\lambda_i - 1)^k \langle t|v_i\rangle\langle v_i|j\rangle\right|^2,$$

*where $\lambda_i$ and $|v_i\rangle$ are eigenvalues and eigenvectors of $P_T$, respectively.*

Proof. (1) **Degree-Normalized Common Neighbors (CN)**: For $k = 2$ and $U_O = I$, the walk operator reduces to $U_W = U_T$. Applying $U_T = 2P_T - I$ twice:

$$U_W^2 = (2P_T - I)^2 = 4P_T^2 - 4P_T + I.$$

The amplitude at node $t$ is:

$$\langle t|U_W^2|j\rangle = 4(P_T^2)_{j,t} - 4(P_T)_{j,t} + \delta_{j,t}.$$

where, $\delta_{j,t}$ is Kronecker delta (1 if $j = t$, 0 otherwise). For $t \neq j$, $\delta_{j,t} = 0$ and $(P_T)_{j,t} = 0$ (assuming no direct edge):

$$\langle t|U_W^2|j\rangle = 4(P_T^2)_{j,t} = 4\sum_k \frac{A_{j,k} A_{k,t}}{\sqrt{d_j d_t} d_k}.$$

Thus, the prediction score is:

$$P(t) \propto \left|4\sum_k \frac{A_{j,k} A_{k,t}}{\sqrt{d_j d_t} d_k}\right|^2 = 16 \left|\sum_k \frac{A_{j,k} A_{k,t}}{\sqrt{d_j d_t} d_k}\right|^2.$$

This matches CN's shared-neighbor count but normalizes contributions by node degrees, suppressing hub bias.

(2) **Resource Allocation (RA)**: For $\tilde{A}_{j,k} = \frac{A_{j,k}}{d_k}$, leading to $\tilde{d}_i = \sum_m \frac{A_{i,m}}{d_m}$. The transition operator is $\tilde{P}_T = \tilde{D}^{-1/2}\tilde{A}\tilde{D}^{-1/2}$. For $U_T = 2\tilde{P}_T - I$, the amplitude after $k = 2$ steps is:

$$\langle t|U_T^2|j\rangle = 4(\tilde{P}_T^2)_{j,t} = 4\sum_k \frac{\tilde{A}_{j,k} \tilde{A}_{k,t}}{\sqrt{\tilde{d}_j \tilde{d}_k \tilde{d}_t}}.$$

Substituting $\tilde{A}_{j,k} = \frac{A_{j,k}}{d_k}$ yields term proportional to:

$$\sum_{k \in \Gamma(j) \cap \Gamma(t)} \frac{1}{d_k^2}.$$

which, after appropriate normalization (and under the assumption that the factor $\sqrt{\tilde{d}_j \tilde{d}_t}$) vary slowly), approximates to:

$$P(t) \propto \left(\sum_{k \in \Gamma(j) \cap \Gamma(t)} \frac{1}{d_k}\right)^2,$$

.

(3) **Adamic-Adar (AA)**: For a reweighted adjacency matrix $\tilde{A}$ with entries $\tilde{A}_{j,k} = \frac{A_{j,k}}{\log d_k}$, where $d_k$ is the original degree of node $k$. The corresponding transition operator can be written as:

$$\tilde{P}_T = \tilde{D}^{-1/2}\tilde{A}\tilde{D}^{-1/2}, \quad \tilde{d}_i = \sum_m \tilde{A}_{i,m}.$$

For $k = 2$ and $U_O = I$, the walk operator becomes $U_W = U_T = 2\tilde{P}_T - I$. Applying $U_T^2$:

$$\langle t|U_T^2|j\rangle = 4(\tilde{P}_T^2)_{j,t} = 4\sum_k \frac{\tilde{A}_{j,k} \tilde{A}_{k,t}}{\sqrt{\tilde{d}_j \tilde{d}_k \tilde{d}_t}}.$$

Substituting $\tilde{A}_{j,k} = \frac{A_{j,k}}{\log d_k}$:

$$P(t) \propto \left| 4 \sum_{k \in \Gamma(j) \cap \Gamma(t)} \frac{1}{\log d_k} \cdot \frac{1}{\sqrt{\tilde{d}_j \tilde{d}_t}} \right|^2.$$

For graphs where $\tilde{d}_i \approx$ const (e.g., slowly varying $\log d_k$), this reduces to:

$$P(t) \propto \left( \sum_{k \in \Gamma(j) \cap \Gamma(t)} \frac{1}{\log d_k} \right)^2,$$

matching the AA heuristic's ordinal ranking.

(4) **Spectral Katz Analogy** : For $k \gg 1$ and $U_O = I$, the walk operator $U_W = U_T$ iteratively applies $U_T = 2P_T - I$. Expanding $U_T$ in its eigenbasis:

$$U_T = \sum_{i=1}^{N} (2\lambda_i - 1)|v_i\rangle\langle v_i|,$$

where $\lambda_i$ and $|v_i\rangle$ are the eigenvalues and eigenvectors of $P_T$. The amplitude at $t$ after $k$ steps is:

$$\langle t|U_W^k|j\rangle = \sum_{i=1}^{N} (2\lambda_i - 1)^k \langle t|v_i\rangle\langle v_i|j\rangle.$$

Substituting $\mu_i = 2\lambda_i - 1$:

$$\langle t|U_W^k|j\rangle = \sum_{i=1}^{N} \mu_i^k \langle t|v_i\rangle\langle v_i|j\rangle.$$

The prediction score is:

$$P(t) \propto \left| \sum_{i=1}^{N} \mu_i^k \langle t|v_i\rangle\langle v_i|j\rangle \right|^2. \quad \mu_i = 2\lambda_i - 1.$$

For graphs with a dominant eigenvalue $\lambda_1 \approx 1$, $\mu_1 \approx 1$, while $|\mu_i| < 1$ for $i \geq 2$. This resembles the Katz Index's damping factor $\beta^l$, but with adaptive spectral damping $\mu_i^k$, ensuring convergence even for irregular graphs.

□

## C.4 Noise Suppression via Interference

**Theorem C.2.** Let $G = (V, E)$ be an undirected graph with adjacency matrix $A$, degree matrix $D$, and normalized adjacency matrix $P_T = D^{-1/2}AD^{-1/2}$. Let $\Delta = 1 - \lambda_2$ denote the spectral gap of $P_T$, where $\lambda_2$ is its second-largest eigenvalue. Let $P_{noise}$ represent topologically irrelevant paths between nodes $j$ and $t$, defined as contributions orthogonal to $P_T$'s principal eigenvector $|v_1\rangle$. Then:

(1) **Quantum Suppression (QSearchNet)**:

$$\sum_{p \in P_{noise}} |QuantumWeight(p)|^2 \leq e^{-4\Delta k},$$

where $k$ is the number of quantum walk steps.

(2) **Classical Damping (Katz Index)**:

$$\sum_{p \in P_{noise}} ClassicalWeight(p) \leq \frac{1 - \Delta}{\Delta}.$$

PROOF. (1) **Quantum Suppression (QSearchNet)**:
From Section C.1, the transition operator $U_T = 2P_T - I$ has eigenvalues $\mu_i = 2\lambda_i - 1$. The spectral gap $\Delta = 1 - \lambda_2 > 0$, guaranteeing non-degenerate eigenvalues and a well-defined separation between the stationary state $|v_1\rangle$ and noise subspace. For noise paths:

$$\mu_i = 2\lambda_i - 1 \leq 2(1 - \Delta) - 1 = 1 - 2\Delta.$$

Let $|j\rangle$ be the initial state localized at node $j$, decomposed as:

$$|j\rangle = \langle v_1|j\rangle|v_1\rangle + \sum_{i=2}^{N} \langle v_i|j\rangle|v_i\rangle,$$

where $\{|v_i\rangle\}$ are orthonormal eigenvectors of $P_T$. We assume the unitary implementation of $U_T$. After $k$ steps of the walk operator $U_W = U_O U_T$, the state evolves as:

$$|\psi_k\rangle = U_W^k|j\rangle.$$

We can define the noise subspace as $\{|v_i\rangle\}_{i \geq 2}^{N}$ and the noise projection $\Pi_{\text{noise}} = I - |v_1\rangle\langle v_1|$. The noise component after $k$ steps is:

$$|\psi_k^{\text{noise}}\rangle = \Pi_{\text{noise}}|\psi_k\rangle.$$

With $U_W$ and the spectral gap $\Delta$, the noise contribution from eigenstate $|v_i\rangle$ ($i \geq 2$) is attenuated by $|\mu_i|^k \leq (1 - 2\Delta)^k$. Thus:

$$\||\psi_k^{\text{noise}}\rangle\| \leq (1 - 2\Delta)^k \||\psi_0^{\text{noise}}\rangle\| \leq (1 - 2\Delta)^k.$$

The noise contribution to $|t\rangle$ is:

$$\langle t|\psi_k^{\text{noise}}\rangle = \sum_{i=2}^{N} \langle t|v_i\rangle\langle v_i|U_W^k|j\rangle.$$

By the Cauchy-Schwarz inequality:

$$\left|\langle t|\psi_k^{\text{noise}}\rangle\right| \leq \sqrt{\sum_{i=2}^{N} |\langle t|v_i\rangle|^2} \cdot \sqrt{\sum_{i=2}^{N} |\langle v_i|U_W^k|j\rangle|^2}.$$

Since $\{|v_i\rangle\}$ forms an orthonormal basis, from Bessel's inequality:

$$\sqrt{\sum_{i=2}^{N} |\langle t|v_i\rangle|^2} \leq 1, \quad \sqrt{\sum_{i=2}^{N} |\langle v_i|U_W^k|j\rangle|^2} \leq (1 - 2\Delta)^k.$$

Thus:

$$\left|\langle t|\psi_k^{\text{noise}}\rangle\right| \leq (1 - 2\Delta)^k.$$

The total noise probability is given by:

$$\sum_{p \in P_{\text{noise}}} |\text{QuantumWeight}(p)|^2 \leq \left((1 - 2\Delta)^k\right)^2.$$

Using the inequality $1 - x \leq e^{-x}$ for $x \in [0, 1]$:

$$(1 - 2\Delta)^{2k} \leq e^{-4\Delta k}.$$

It is important to note that $U_O$ mixes the eigenstates of $U_T$ after each step. As $U_O$ is unitary, it preserves the overall probability of the subspace. Thus, the above remains valid as a worst-case scenario, though, this mixing may lead to faster convergence in some cases.

(2) **Classical Damping (Katz Index)**:

The Katz index assigns a weights to paths of length $l$ as $\alpha^l$, where $\alpha \in (0, 1)$ ensures convergence of infinite sum. The total noise contribution from all the paths can be expressed as:

$$\sum_{p \in \mathcal{P}_{\text{noise}}} \text{ClassicalWeight}(p) = \sum_{l=1}^{\infty} \alpha^l \langle t| P_T^l |j\rangle_{\text{noise}}.$$

As $P_T = \sum_{i=1}^{N} \lambda_i |v_i\rangle \langle v_i|$, the noise term (i.e, $i \geq 2$) becomes:

$$\langle t| P_T^l |j\rangle_{\text{noise}} = \sum_{i=2}^{N} \lambda_i^l \langle t|v_i\rangle\langle v_i|j\rangle.$$

Interchanging the summation over path lengths $l$ with the sum over noise eigenstates, we have:

$$\sum_{p \in \mathcal{P}_{\text{noise}}} \text{ClassicalWeight}(p) = \sum_{i=2}^{N} \langle t|v_i\rangle\langle v_i|j\rangle \sum_{l=1}^{\infty} (\alpha \lambda_i)^l.$$

For $\alpha \in (0, 1)$, the geometric series converges as:

$$\sum_{l=1}^{\infty} (\alpha \lambda_i)^l = \frac{\alpha \lambda_i}{1 - \alpha \lambda_i}.$$

To maximize the bound while ensuring convergence, let $\alpha \to 1^-$ (approaching from below). Parameterize $\alpha = 1 - \epsilon$ where $\epsilon \to 0^+$. For $i \geq 2$, we have $\lambda_i \leq 1 - \Delta$, giving:

$$\frac{\alpha \lambda_i}{1 - \alpha \lambda_i} \leq \frac{(1-\epsilon)(1-\Delta)}{1 - (1-\epsilon)(1-\Delta)}.$$

Simplify the denominator as: $1 - (1-\epsilon)(1-\Delta) = \Delta + \epsilon(1-\Delta)$.. We can write:

$$\frac{(1-\epsilon)(1-\Delta)}{\Delta + \epsilon(1-\Delta)}.$$

Taking the limit $\epsilon \to 0^+$:

$$\lim_{\epsilon \to 0^+} \frac{(1-\Delta) - \epsilon(1-\Delta)}{\Delta + \epsilon(1-\Delta)} = \frac{1-\Delta}{\Delta}.$$

Apply the Cauchy-Schwarz inequality to the eigenvector overlaps:

$$\sum_{i=2}^{N} |\langle t|v_i\rangle\langle v_i|j\rangle| \leq \sqrt{\sum_{i=2}^{N} |\langle t|v_i\rangle|^2} \sqrt{\sum_{i=2}^{N} |\langle v_i|j\rangle|^2} \leq 1.$$

Summing over all $i \geq 2$:

$$\sum_{p \in \mathcal{P}_{\text{noise}}} \text{ClassicalWeight}(p) \leq \frac{1-\Delta}{\Delta}.$$

Quantum noise decays as $O(e^{-4\Delta k})$, while classical noise saturates at $\frac{1-\Delta}{\Delta}$. The suppression ratio:

$$\frac{\text{Quantum Noise}}{\text{Classical Noise}} \leq \frac{e^{-4\Delta k}}{(1-\Delta)/\Delta} \to 0 \quad \text{as } k \to \infty.$$

□

## D Simulation on a Classical Computer

In this section, we describe the classical simulation methodology used to calculate link prediction scores on a conventional computer. The simulation approximates the dynamics of quantum walk with sparse matrix operations, implemented via a COO tensor, to mimic amplitude redistribution and amplification. The transition subspace ($P_T$) can be represented as:

$$A^{\text{normalized}}(i, j) \propto \frac{A(i, j)}{\sqrt{d(i) \cdot d(j)}} \qquad (22)$$

where, $A$ is the adjacency matrix capturing the connections in the graph, and $D$ is the diagonal degree matrix, with $D_{(i,i)} = d(i)$ representing the degree of node $i$. The oracle phase shift operator ($U_O$) is classically implemented by applying a sign flip to the amplitude corresponding to the target node $t$. The quantum walk evolves iteratively:

$$\psi(t+1) = U_O \cdot U_T \cdot \psi(t) \qquad (23)$$

Here, $U_O$ and $U_T$ act on the state vector $\psi(t)$ to update the walker's amplitude distribution. At the end of the simulation, the probability of the system being at the target node $t$ is calculated by taking the squared magnitude of the corresponding amplitude:

$$P(t) = |\psi(t)|^2$$

This final step simulates the probabilistic nature of quantum measurement where $P(t)$ represents the likelihood of a link between the initial and target nodes based on simulated quantum walk dynamics.